\documentclass[twocolumn,twoside,slac_two]{revtex4}
\usepackage{graphicx}
\usepackage{fancyhdr}

\begin{document}

\title{Tracking Down the Highest Spindown Power Gamma-ray Pulsars}

%

\author{X. Hou, D. Dumora, D.A. Smith, M. Lemoine-Goumard}
\affiliation{Universit\'e Bordeaux 1, CNRS/IN2P3, Centre d'Etudes Nucleaires de Bordeaux-Gradignan, France}
\author{M.-H.Grondin}
\affiliation{Max-Planck-Institut f\"ur Kernphysik, Saupfercheckweg 1, 69117 Heidelberg, Germany}
\author{For the \emph{Fermi} Large Area Telescope Collaboration and the \emph{Fermi} Pulsar Timing Consortium \cite{psc}}

\begin{abstract}
Forty six $\gamma$-ray pulsars were reported in the First \emph{Fermi} Large Area Telescope (LAT) Catalog of Gamma-ray Pulsars \cite{1stcat}. Over forty more have been seen since then. A simple but effective figure-of-merit for $\gamma$-detectability is $\sqrt{\dot E}/d^2$, where $\dot E$ is the pulsar spindown power and $d$ the distance. We are tracking down the best $\gamma$-ray candidates not yet seen. We present the timing and spectral analysis results of some new high spindown power, nearby $\gamma$-ray pulsars. We also update some population distribution plots in preparation for the 2nd \emph{Fermi} LAT $\gamma$-ray Pulsar Catalog.

\end{abstract}

\maketitle

\thispagestyle{fancy}


\section{Introduction}
The launch of the \emph{Fermi} Gamma-ray Space Telescope has opened a new era in the study of $\gamma$-ray pulsars. Besides the 7 previously-known CGRO $\gamma$-ray pulsars, \emph{Fermi} Large Area Telescope (LAT) has detected 88 $\gamma$-ray pulsars, among them 46 reported in the First \emph{Fermi} Large Area Telescope (LAT) Catalog of Gamma-ray Pulsars \cite{1stcat}. These $\gamma$-ray pulsars fall into three different classes: young or middle-aged radio selected (using radio ephemerides), young or middle-aged gamma selected (by blind periodicity searches), and millisecond pulsars (MSP, currently only radio selected). Many of the radio MSPs were recently discovered at the positions of unidentified \emph{Fermi} sources. The increasing population of $\gamma$-ray pulsars enables great improvements in the understanding of the $\gamma$-ray emission mechanisms and the Galactic population of pulsars through detailed study of their light curves and spectra. We present here the timing and spectral analysis results for some of these new high spindown power, nearby $\gamma$-ray pulsars, as well as some preliminary population distribution plots for the 2nd \emph{Fermi} LAT $\gamma$-ray Pulsar Catalog.

\section{New detections of young energetic $\gamma$-ray pulsars}
Figure 1 shows (colored points) the 88 $\gamma$-ray pulsars (including the 7 previously-known CGRO $\gamma$-ray pulsars) detected with the \emph{Fermi} LAT to date. 55 were discovered using radio ephemerides and 26 by blind period searches. We are tracking down the best $\gamma$-ray candidates not yet seen.

\begin{figure*}[!htp]
\includegraphics[scale=0.45]{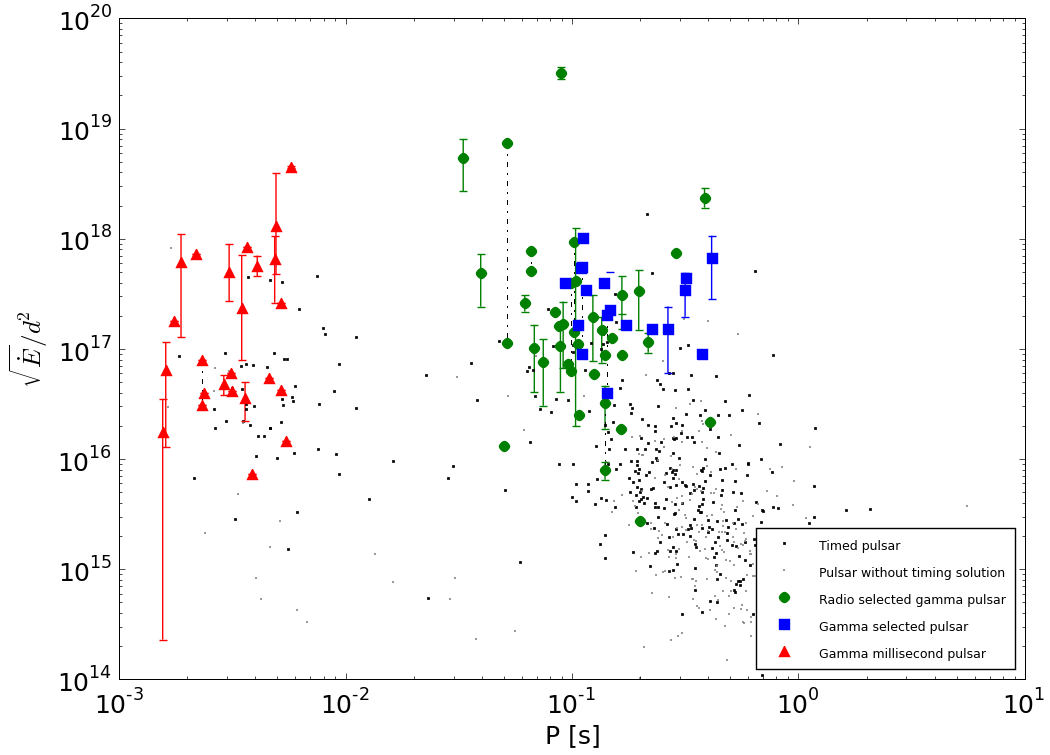}
\caption{Pulsar $\gamma$-detectability figure-of-merit $\sqrt{\dot E}/d^2$ vs spin period. PRELIMINARY} 
\end{figure*}

Ranked by $\sqrt{\dot E}/d^2$ for ATNF \cite{Atnf} pulsars with $\dot E >\times 10^{33}$ erg s$^{-1}$ \& $P_{0} > 10$ ms, nine recently-detected young energetic radio-selected $\gamma$-ray pulsars are listed in Table 1 along with the preliminary timing analysis results.

\begin{figure}[!htp]
\centering
\includegraphics[scale=0.41]{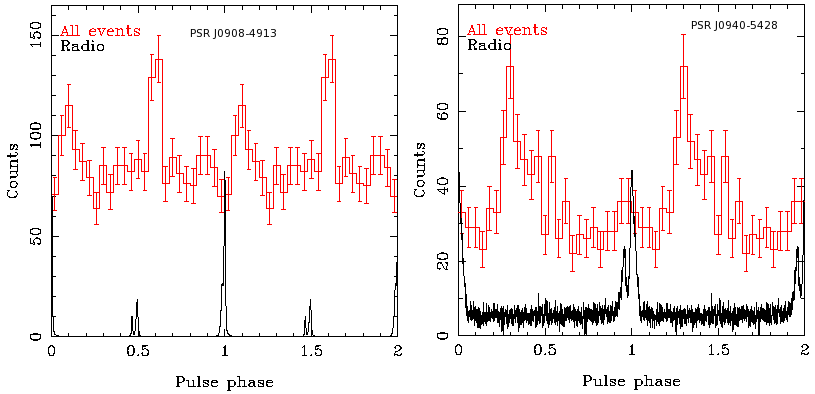}
\caption{Multiple energy bands $\gamma$-ray light curves superimposed with radio profile. PRELIMINARY} 
\end{figure}

Figure 2 shows the preliminary $\gamma$-ray and radio profiles for two new $\gamma$-ray pulsars, J0908-4913 and J0940-5428. They have either two narrow peaks or one single wide pulse, typical for $\gamma$-ray pulsars. 

\begin{figure}[!htp]
\centering
\includegraphics[scale=0.33]{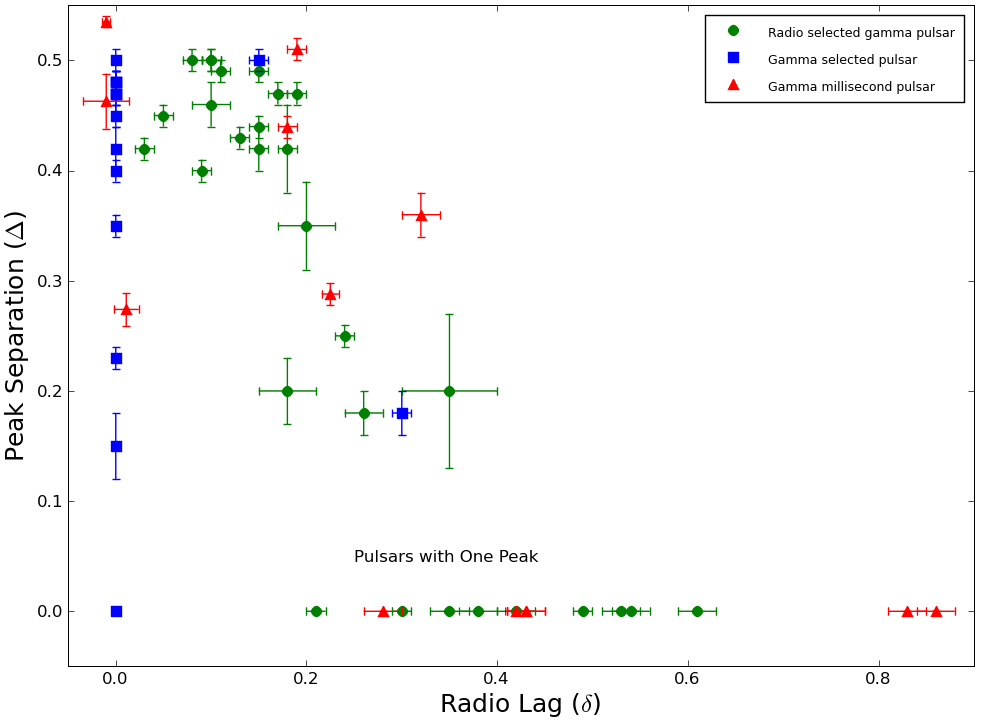}
\caption{$\Delta$-$\delta$ distribution for 88 $\gamma$-ray pulsars detected by \emph{Fermi} Large Area Telescope (LAT). PRELIMINARY } 
\end{figure}

The gamma peak separation - radio lag distribution ($\Delta$-$\delta$ distribution, Figure 3) can be compared to predictions of the geometrical models of outer magnetosphere emission, e.g. ``Two Pole Caustic" (TPC) \cite{TPC} and ``Outer Gap" (OG) \cite{OG}, and help us probe details of the emission geometry. In particular, PSR J0908-4913 has been determined to be an orthogonal rotator through geometry studies in the radio domain \cite{J0908}. 

Why are some pulsars invisible in $\gamma$-rays?  As discussed by Dumora et al. \cite{sardinia}:
\begin{itemize}
\item Distance \& the Galactic diffuse $\gamma$-ray background influence the signal-to-noise ratio. We see PSR J1410-6132 in spite of a very  low $\sqrt{\dot E}/d^2$. Its large DM distance bears further investigation.  
\item Timing model quality. PSR J1357-6429 was discovered once we obtained a good radio ephemeris.
\item Physics (beam geometry): for those pulsars having reliable distance measurements and good timing models, exploration of geometry will help us constrain the emission models and improve the population syntheses \cite{sublum}.
\end{itemize}

\section{Zoom on PSR J1357-6429}
Discovered during the Parkes multibeam survey of the Galactic plane \cite{camilo04}, PSR J1357-6429 is among the youngest and most energetic nearby pulsars known. It was in the top 2 “highest $\sqrt{\dot E}/d^2$ pulsars not yet seen with the LAT” \cite{sardinia} and might have become a candidate ``gamma faint pulsar" \cite{sublum}. In fact, the absence of $\gamma$-ray pulsations was due to the lack of a good radio timing model \cite{lemoine11}. We present the LAT analysis result using 29 months of data covering Aug 4, 2008 to Jan 15, 2011. We selected events with energies greater than 0.1 GeV but excluded those with zenith angles  larger than 100$^{\circ}$ to minimize contamination from secondary $\gamma$-rays from the Earth’s atmosphere. We used Source events class which optimizes the trade-off between $\gamma$-ray detection efficiency and residual charged-particle contamination in a manner well-suited to localized, persistent sources, as distinct from the Transient or Diffuse classes. The events were analyzed using the standard software package \emph{ScienceTools} \cite{FSSC}, and photon phases were calculated using the ``Fermi-plugin" \cite{plugin} available in the TEMPO2 \cite{tempo2} pulsar timing software. We used the most recent P7SOURCE\_V6 instrument response function (IRFs) as well as the Galactic Diffuse Model: \emph{gal\_2yearp7v6\_v0.ﬁts} and the Extra-galactic Diffuse Model: \emph{iso\_p7v6source.txt} as documented at the Fermi Science Support Center \cite{FSSC}.

\begin{figure}[!ht]
\centering
\includegraphics[scale=0.35]{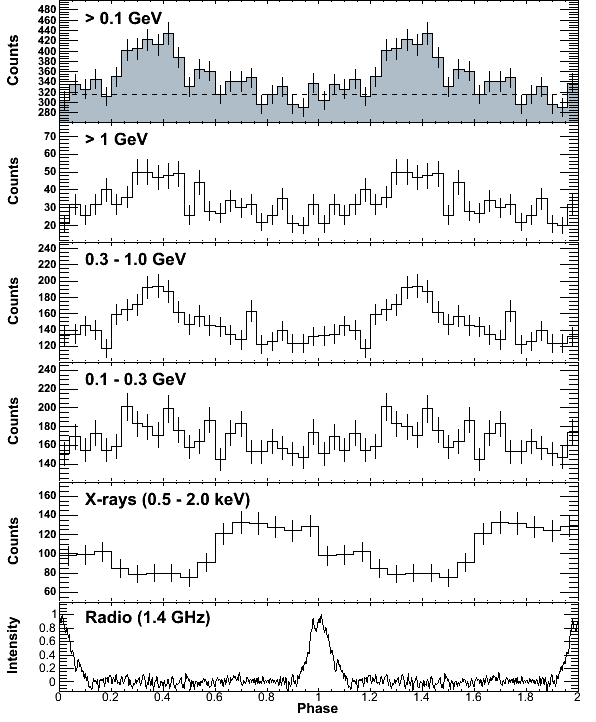}
\caption{Top three panels: phase-aligned $\gamma$-ray histograms of PSR J1357-6429 within an energy-dependent circular region of $0.8^{\circ}$ around the radio pulsar position. Two rotations are plotted with 25 bins per period. The dashed line shows the background level estimated using two nearby circular regions of 1.5$^{\circ}$ from the pulsar. Second panel from bottom: X-ray pulse profile extracted from the XMM-Newton data of 2009 in the 0.5-2 keV energy band. Two rotations are plotted with 15 bins per period. Bottom panel: radio pulse profile based on Parkes observations at a center frequency of 1.4 GHz with 256 phase bins. \cite{lemoine11}} 
\end{figure}

\subsection{Timing analysis}
Figure 4 shows the $\gamma$-ray light curves at energy above 0.1 GeV and within $0.8^{\circ}$ around the radio pulsar position using an energy--dependent cone of radius $\theta _{68}\le \sqrt{(5.3^{\circ} \times (\frac{E}{100\text{Mev}})^{-0.745})^2+0.09^2}$ . This choice takes into account the instrument performance to maximize the signal-to-noise ratio over a broad energy range. Performing an unbinned maximum-likelihood fit of the phases to a constant plus single Gaussian, we obtained an intrinsic pulsed fraction of 0.56 $\pm$0.13 with a Gaussian of width 0.48$\pm$0.18, centered at phase 0.83$\pm$0.03. X-ray (XMM-Newton 2009 data) and radio (1.4GHz from Parkes \cite{radio}) profiles are also presented, confirming of X-ray pulsations first suggested by Zavlin (2007).

\subsection{Spectral analysis}
The phase-averaged spectrum of PSR J1357-6429 was obtained using \emph{gtlike}, a maximum likelihood spectral analysis \cite{gtlike} implemented in the \emph{Fermi ScienceTools}. The spectrum is best described by a power~law with a spectral index of $\Gamma=1.5\pm0.3\pm0.3$ with an exponential cut~off at $E_{c}=0.8\pm0.3\pm 0.3$ GeV and an integral photon flux above 100 MeV of $F_{100}=(6.5\pm1.6\pm2.3)\times 10^{-8}$ ph cm$^{-2}$ s$^{-1}$. The integral energy flux above 100 MeV is $F_{100}=(1.9\pm0.2\pm0.8)\times 10^{-5}$ MeV cm$^{-2}$ s$^{-1}$. Assuming a distance of 2.4 kpc, the gamma~ray luminosity is $L_{\gamma} = (2.1\pm0.3\pm0.8) \times10^{34}$ erg s$^{-1}$, consistent with a  $L_{\gamma}\propto \sqrt{\dot E}$ relationship. The first error is statistical, while the second represents our estimate of systematic effects. Figure 5 shows the best fit spectrum as well as the $95\%$ C.L. upper~limits on the $\gamma$-ray emission from its potential Pulsar Wind Nebula HESS J1357-645 \cite{lemoine11}.

\begin{figure}[!ht]
\centering
\includegraphics[scale=0.42]{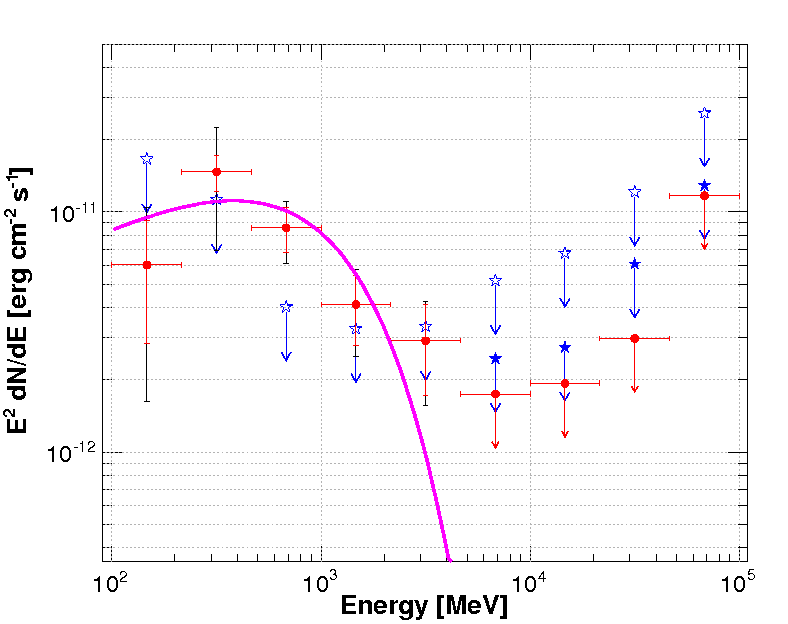}
\caption{Spectral energy distribution of PSR J1357-6449 in gamma rays with LAT. Red dots are from the maximum-likelihood fit in individual energy bands between 100 MeV and 100 GeV with the statistical significance $\ge$3$\sigma$, otherwise an upper limit arrow is shown. The statistical errors are shown in red and in black both the statistical and systematic errors summed in quadrature. The magenta solid line reperesents the maximum-likelihood fit by a power law with an exponentially cut-off. Blue filled (open) stars represent the $95\%$ C.L. upper limits derived assuming a Gaussian of 0.2$^{\circ}$ at the position of HESS J1356-645 in the whole signal (in the off-pulse). \cite{lemoine11}}
\end{figure}

\section{Conclusion}
The number of $\gamma$-ray pulsars has increased steadily since the launch of the \emph{Fermi} Satellite in June 2008 and has reached 88 including the most recent detection of J0940-5428, the last piano key missing. We report 9 recently-detected young energetic radio-selected $\gamma$-ray pulsars ranked by a figure-of-merit for $\gamma$-detectability with some light curves and a preliminary update of the population distribution plots. The quality of timing models can influence or dominate $\gamma$-ray detection. The incoming 2nd \emph{Fermi} LAT $\gamma$-ray Pulsar Catalog will provide a larger population which will give an opportunity to better constrain the $\gamma$-ray emission models. 

\begin{table*}[!ht]
\caption{Nine recently-detected young energetic radio-selected $\gamma$-ray pulsars}
\begin{center}

\begin{tabular}{|l|c|c|c|c|c|c|c|c|c|}
\hline \text{Rank} & \text{PSRJ} & \text{$P_{0}$} & \text{$\dot E$}
& \text{$\sqrt{\dot E}/d^2$} & \text{Dist1} & \text{Gb} & \text{Npeak}
& \text{radio lag} & \text{$\gamma$-ray peak sep.} \\
& &(s) & (erg s$^{-1}$) & (erg s$^{-1}$cm$^{-2}$) &(kpc) & (deg) & & $\delta$ & $\Delta$\\
\hline 15 & J1357-6429 & 0.1661 & 3.10E+36 & 2.82E+17 & 2.50 & -2.51 & 1 & 0.38$\pm$0.02 &...\\
\hline 18 & J1531-5610 & 0.0842 & 9.09E+35 & 2.18E+17 & 2.09 & 0.03 & 1 & 0.35$\pm$0.02 &...\\
\hline 23 & J0940-5428 & 0.0875 & 1.93E+36 & 1.58E+17 & 2.95 & -1.29 & 1 & 0.30$\pm$0.01 &...\\
\hline 28 & J0908-4913 & 0.1068 & 4.92E+35 & 1.09E+17 & 2.53 & -1.01 & 2 & 0.10$\pm$0.01 & 0.50$\pm$ 0.01 \\
\hline 37 & J1730-3350 & 0.1395 & 1.23E+36 & 8.74E+16 & 3.54 & 0.09 & 2 & 0.05$\pm$0.01 & 0.45$\pm$ 0.01 \\
\hline 51 & J1801-2451 & 0.1249 & 2.59E+36 & 5.92E+16 & 5.22 & -0.88 & 2 & 0.10$\pm$0.02 & 0.46$\pm$ 0.02 \\
\hline 87 & J1016-5857 & 0.1074 & 2.58E+36 & 2.52E+16 & 8.0 & -1.88 & 2 & 0.10$\pm$0.01 & 0.50$\pm$ 0.01 \\
\hline 105 & J1648-4611 & 0.1650 & 2.09E+35 & 1.86E+16 & 4.96 & -0.79 & 1 & 0.53$\pm$0.01 &... \\
\hline 129 & J1410-6132 & 0.0500 & 1.01E+37 & 1.30E+16 & 15.6 & -0.09 & 2 & 0.03$\pm$0.01 & 0.42$\pm$ 0.01 \\
\hline
\end{tabular}
\label{table1}
\end{center}
\end{table*}

\newpage
\bigskip 

\end{document}